\documentclass[preprint,pre,aps]{revtex4}
\usepackage{epsfig}
\usepackage{amsmath}
\usepackage{amsfonts}
\usepackage{oldgerm}

\begin{document}
\preprint{}

\title{DEFLECTIONS IN MAGNET FRINGE FIELDS}
\author{Y.~Papaphilippou}\altaffiliation{Present address: European
Synchrotron Radiation Facility, BP 220, F-38043 Grenoble Cedex,
FRANCE}\email{yannis@esrf.fr}  \affiliation{Brookhaven National
Laboratory, Upton, New York 11973,   U.S.A.}  \author{J.~Wei}
\affiliation{Brookhaven National Laboratory, Upton, New York 11973,
U.S.A.}  \author{R.~Talman}  \affiliation{Cornell University, Ithaca,
New York 14843,   U.S.A.}  \date{January 30, 2003}

\begin{abstract}
A transverse multipole expansion is derived, including the
longitudinal components necessarily present in regions of varying
magnetic field profile. It can be used for exact numerical orbit
following through the fringe field regions of magnets whose end
designs introduce no extraneous components, {\it i.e.} fields not
required to be present by Maxwell's equations. Analytic evaluations of
the deflections are obtained in various approximations.  Mainly
emphasized is a ``straight-line approximation'',  in which particle
orbits are treated as straight lines through  the fringe field
regions. This approximation leads to a readily-evaluated figure of
merit, the ratio of r.m.s. end deflection to nominal body deflection,
that can be used to determine whether or not a fringe field can be
neglected.  Deflections in ``critical'' cases  (e.g. near intersection
regions) are analysed in the same approximation.
\end{abstract}

\pacs{PACS numbers:}

\maketitle


\section{STRATEGY AND NOTATION}

The purpose of this paper is to derive formulas for the orbit
deflections caused by the fringe fields of non-solenoidal accelerator
magnets. The main ingredient is a multipole expansion for fields
having arbitrary longitudinal profile and including all field
components (and only those) required to be present by Maxwell's
equations.

Because terminology describing magnets depends on context, we define
some of our terms, if only implicitly, by using them in this section.
Most magnets in accelerators are ``dipoles'', ``quadrupoles'' or other
``multipoles'' where, in this paper, we distinguish by quotation marks
the common names of these magnets from the dipole, quadrupole,
multipole, etc., terms appearing in mathematical expansions of their
magnetic fields. The particle orbits are {\it paraxial}, with {\it
small} transverse displacements, $r=(x^2+y^2)^{1/2}$, with slopes
$(x',y')\equiv(dx/dz,dy/dz)$ small compared to 1 because the orbits
are more or less parallel to the $z$-axis, which is the magnet
centerline.  The dominant magnetic field components ($B_x,B_y$) are
therefore {\it transverse} to this axis, and the currents in most
accelerator magnets are therefore {\it longitudinal}.  But actual
magnet coils must have radial leads to return the currents and,
because of practical considerations, they also have azimuthal currents.

The standard multipole expansion derives entirely from longitudinal
magnet currents (this includes the bound currents in ferromagnets).
It is only  for a {\it long} magnet whose length $L$ is large (for
example compared to a typical radial magnetic half-aperture $r_{1/2}$)
that a single multipole term provides a good approximation to the
field. Yet, as concerns the effect of the magnet on a particle orbit,
a common idealization is the {\it short magnet} or {\it thin lens}
approximation, in which the entire deflection caused by the magnet
occurs at a single longitudinal position. Even more extreme than our
straight line approximation is to treat the transverse orbit
coordinates $(x,y)$ as constant through the entire magnet, body and
ends; the deflection (say horizontal) is proportional to a {\it field
integral} of the form $\Delta
x'(x,y)\sim\int_{-\infty}^{\infty}\,{\cal B}(x,y,z)\,dz$, where ${\cal
B}(x,y,z)$ stands for any one of $B_x,B_y,dB_x/dx,dB_x/dy,\dots$, that
is, either of the transverse magnetic field components, or any of
their derivatives with respect to $x$ and/or $y$.  Commonly then, one
defines an {\it effective magnet length} $L_{\rm eff}\approx L$ such
that  $\int_{-\infty}^{\infty}\,{\cal B}(0,0,z)\,dz={\cal
B}(0,0,0)\,L_{\rm eff}$.  This length is specific to the particular
multipole the magnet is  designed to produce.  In spite of the facts
that the magnet must be long to validate the multipole approximation,
yet short to validate the thin element treatment, and that
discontinuous magnetic fields violate the Maxwell's equations, this
approximation is curiously accurate for most accelerator magnets.
Because of this good start, it promises to be effective to improve
upon the approximation by assuming that magnets have ideal multipole
fields within the length $L_{\rm eff}$, but  also to include ``end
fields'' applicable in regions of length $\Delta L_-$ and $\Delta L_+$
at input and output ends. In this approximation the transverse
magnetic fields are continuous, but their derivatives are
discontinuous at both ends of the fringe field regions.

In a well-designed magnet, the same multipole that is dominant in the
central region is dominant in the end regions. But the fields in the
end regions are necessarily more complicated and include longitudinal
components $B_z(x,y,z)$. Since the fields in these regions are, in
principle, constrained only by Maxwell equations, rigorous formulas
for the deflections they cause can only be evaluated by solving
differential equations appropriate for the detailed magnet end
configuration. To obtain analytic formulas we must make some
assumptions, the first of which is that the formulation is not
intended to apply to ``intentional solenoids'' (because of their large
azimuthal currents and longitudinal field components). Furthermore the
only longitudinal fields included are those that are required by
Maxwell's equation to be present in regions of varying longitudinal
profile. In other words, the formulas can be expected to be accurate
for ``well-designed'' magnets, in which the dominant fringe field
multipolarity matches the body multipolarity. This can, in principle,
be assured by proper shaping of pole ends and proper conformation of
the magnet return currents. In the absence of magnetic field
measurements in the end regions, this is the only practical assumption
one can make when predicting the fringe field deflections. If the
fields {\it have\/} been accurately measured or calculated, to improve
on formulas given in this paper, it would be necessary to separate out
the (presumably small) extraneous components and include their effects
perturbatively. One cannot exclude the possibility of end  geometries
that introduce multipoles for which the extraneous fringe fields are
large compared to the required fringe fields,  either intentionally of
unintentionally. The present formalism would not be directly
applicable for such fields.

In this paper, we derive first approximations for the deflections
occurring in the end field regions, of the form $\Delta
x'_-\sim\int_{-\Delta L_-}^0\,{\cal B}(x,y,z)\,dz$ and $\Delta
x'_+\sim\int_L^{L+\Delta L_+}\,{\cal B}(x,y,z)\,dz$~\footnote{In most
cases, one limit of each integral can be taken to infinity, because of
the rapid fall off of the field.}.  Like the thin lens approximation,
these formulas assume the transverse orbit displacement is constant
through the end intervals $\Delta L_-$ and $\Delta L_-$. This is a
much more valid assumption than assuming constant displacement through
the whole magnet if, as is usually true, the end regions are
``short''; $\Delta L_{\pm}<<L$. Furthermore terms proportional  to
transverse slopes $x'$ and $y'$ can be consistently included, in the
formulas for the deflections.

A criterion for the validity of treating the end region as short can
be based on the inequality $|\beta'_{x,y}|\Delta
L_{\pm}/\beta_{x,y}<<1.$, where $\beta_{x,y}$ and $\beta'_{x,y}$ are
the usual beta functions  and their derivatives with respect to the
longitudinal position $z$.  When this is true the (fractional) rate of
change of multipole strength $1/\Delta L_{\pm}$ is large compared to
the (fractional) rate of change of lattice beta functions.

There is often a tendency  to believe that multipole contributions
from opposite ends of a magnet cancel each other. But, since this is
not universally valid, in this paper no such assumption will be made.

\section{3D MULTIPOLE EXPANSION}

In this section, a multipole expansion is developed that is
appropriate for performing the calculation just described.   This
expansion is applicable to magnetic fields that depend arbitrarily on
the longitudinal coordinate $z$ but, being a power series in the
transverse coordinates $x$ and $y$, its accuracy after truncation to
an order $n$ deteriorates at large transverse amplitudes.  The
expansion is intended to describe an arbitrary ``multipole'' magnet
along with its fringe field.  The formalism presented here generalizes
an approach described by Steffen and reduces to formulas he gives in
the case of ``dipoles'' and ``quadrupoles''~\cite{steffen}.

In the current-free regions to which the beams are restricted, the
magnetostatic field ${\bf B}(x,y,z)$ can be expressed as the gradient
of a scalar potential $\Phi(x,y,z)$
\begin{equation}
{\bf B}(x,y,z)={\bf \nabla} \Phi(x,y,z) = \frac{\partial
\Phi}{\partial x}{\bf x} +\frac{\partial \Phi}{\partial y}{\bf y}+
\frac{\partial \Phi}{\partial z}{\bf z} \;\;,
\label{eq:fieldgrad}
\end{equation}
where $\Phi$ satisfies
\begin{equation}
\nabla^2 \Phi(x,y,z)= \frac{\partial^2 \Phi}{\partial x^2} +
\frac{\partial^2 \Phi}{\partial y^2} + \frac{\partial^2 \Phi}{\partial
z^2} = 0 \;\;.
\label{eq:laplace}
\end{equation}
An appropriate expansion is
\begin{equation}
  \Phi(x,y,z)= \sum_{m=0}^\infty\sum_{n=0}^\infty {\cal C}_{m,n}(z)
  \frac{x^n\, y^m}{n!\, m!}\;\;,
\label{eq:potential}
\end{equation}
where the coefficients ${\cal C}_{m,n}(z)$ depend on the longitudinal
position $z$\footnote{The spatial dependence of function $\Phi$ can
guide the shaping of the pole pieces of iron magnets to match, as
closely as possible, equipotentials of $\Phi$. This is discussed by
Steffen~\cite{steffen} for the case of quadrupoles.}.

Substituting Eq.~\eqref{eq:potential} into Eq.~\eqref{eq:laplace}, we
get a recursion relation for the coefficients;
\begin{equation}
  {\cal C}_{m+2,n} = - {\cal C}_{m,n+2} - {\cal C}^{[2]}_{m,n}\;\;,
\label{eq:coef}
\end{equation}
where in this and subsequent formulas a superscript $[l]$ denotes $l$
differentiations with respect to $z$; in this case $l=2$.  Now, we can
evaluate the gradient of the potential and get the field components in
the three Cartesian directions
\begin{equation}
\begin{split}
B_x(x,y,z) & = \sum_{m=0}^\infty\sum_{n=0}^\infty {\cal C}_{m,n+1}(z)
  \frac{x^n\, y^m}{n!\,m!} \\ B_y(x,y,z) & =
  \sum_{m=0}^\infty\sum_{n=0}^\infty {\cal C}_{m+1,n}(z) \frac{x^n\,
  y^m}{n!\,m!}  \\ B_z(x,y,z) & = \sum_{m=0}^\infty\sum_{n=0}^\infty
  {\cal C}^{[1]}_{m,n}(z) \frac{x^n\, y^m}{n!\,m!}
\end{split} \;\;.
\label{eq:field}
\end{equation}
The two-index coefficients ${\cal C}_{m,n}$ can be expressed in terms
of the usual normal and skew multipole coefficients which, as well as
being conventional, have only one index,
\begin{equation}
\begin{split}
b_n(z) = &{\cal C}_{1,n}(z)  = \left ( \frac{\partial^n B_y}{\partial
x^n}\right){\bigg|}_{x=y=0} (z)   \\ a_n(z) = &{\cal C}_{0,n+1}(z) =
\left ( \frac{\partial^n B_x}{\partial x^n}\right){\bigg|}_{x=y=0} (z)
\end{split} \;\;.
\label{eq:mult}
\end{equation}
We next seek a representation of the field as a function of these
coefficients and their derivatives. The relation~\eqref{eq:coef} can
be applied recursively to obtain
\begin{equation}
  {\cal C}_{m,n} = \sum_{l=0}^{k} (-1)^k\binom{k}{l} {\cal
  C}^{[2l]}_{m-2k,n+2k-2l}  \;\;,
\label{eq:rec1}
\end{equation}
where the upper limit of the series $k$ is equal to the integer part
of $m/2$. This shows that the coefficients ${\cal C}_{m,n}$ can be
expressed as  a series of even derivatives of  ${\cal C}_{0,n+1}$ or
${\cal C}_{1,n}$.  Using Eq.~\eqref{eq:mult} we can distinguish two
cases for $m$, namely $m=2k$ (even) or $m=2k+1$ (odd), and we have
\begin{equation}
\begin{split}
{\cal C}_{0,0}=0,\quad{\cal C}_{2k,n} & =  \sum_{l=0}^{k} (-1)^{k}
\binom{k}{l} a^{[2l]}_{n+2k-2l-1}, \quad\hbox{for\ $n>0$},
\\ {\cal C}_{2k+1,n}  & =  \sum_{l=0}^{k} (-1)^{k} \binom{k}{l}
b^{[2l]}_{n+2k-2l}
\label{eq:rec2}
\end{split} 
\;\;.
\end{equation}
The requirement ${\cal C}_{0,0}=0$ corresponds to the restriction to
non-solenoidal magnets.

Substituting this representation into Eqs.~\eqref{eq:field} and
rearranging the $m$-summation yields
\begin{equation}
\begin{split}
 B_x(x,y,z) = &\sum_{n=0}^\infty\sum_{m=0}^\infty\sum_{l=0}^{m} (-1)^m
  \binom{m}{l} \frac{x^n\, y^{2m}}{n!\,(2m)!} \left(
  b^{[2l]}_{n+2m+1-2l}{\frac{y}{2m+1}} + a^{[2l]}_{n+2m-2l} \right)  \\
B_y(x,y,z)  =& \sum_{n=0}^\infty\sum_{m=0}^\infty (-1)^m
\frac{x^n\,y^{2m}}{n!\,(2m)!}  \Biggl[\sum_{l=0}^{m}\binom{m}{l}
b^{[2l]}_{n+2m-2l} \\ &  \qquad\qquad\qquad\qquad\quad
-\sum_{l=0}^{m+1} \binom{m+1}{l} a^{[2l]}_{n+2m+1-2l}{\frac{y}{2m+1}}
\Biggr]  \\
 B_z(x,y,z)  = &\sum_{n=0}^\infty\sum_{m=0}^\infty\sum_{l=0}^{m}
 (-1)^m \binom{m}{l}  \frac{x^n\, y^{2m}}{n!\,(2m)!} \left(
 b^{[2l+1]}_{n+2m-2l}{\frac{y}{2m+1}} + a^{[2l+1]}_{n+2m-1-2l} \right)
\end{split} \;\;,
\label{eq:fieldmgen} 
\end{equation}
again limiting the ranges so the lowest coefficients are $b_0\equiv
C_{1,0}$ and $a_0\equiv C_{0,1}$.

In an idealized model of a magnet, only one (or in the case of
combined function magnets, two) of the multipole coefficients will be
non-vanishing in the body of the magnet (length $L_{\rm eff}$) and in
this region only the $l=0$ terms in the expansions survive.  The
important terms are: $(m=0,l=0)$ corresponding to the leading
``design'' multipole; $(m=0,l=1)$, the ``next-to-leading'' term
associated with longitudinal variation of the design multipole; and
$(m=1,l=0)$ coming from the next higher body multipole. Examples in
this paper are mainly concerned with the relative importance of the
first two of these terms in the deflections caused by the actual
magnet, including body and ends. The same formulas could, however, be
used to evaluate the relative importance of the second and third
terms---to answer the question ``Which are more important, fringe
fields or body field imperfection?''

To obtain results concerning the symmetries of the skew and normal
multipole coefficients it is more useful to express these formulas in
terms of cylindrical coordinates. This is done in Appendix~A.

In the fringe regions of the magnet, the fields can be arranged so
that they match the central fields at the ends of the body region and
fall linearly to zero in the fringe regions. For example, let us keep
just one more term as a ``next approximation'', arrange  its leading
($l=0$) part to match a given body field at $z=0$, and let it vary
linearly with $z$;
\begin{equation}
\begin{split}
& B_x(x,y,z)  \approx \sum_{n=1}^\infty\sum_{m=0}^\infty
  \frac{x^{n-1}\, y^{2m+1}}{(n-1)!\,(2m+1)!}(-1)^m \left[
  b^{[0]}_{n+2m} + b^{[1]}_{n+2m}\,z \right] \\ & B_y(x,y,z)  \approx
  \sum_{n=0}^\infty\sum_{m=0}^\infty \frac{x^n\,
  y^{2m}}{n!\,(2m)!}(-1)^m  \left[  b^{[0]}_{n+2m} + b^{[1]}_{n+2m}\,z
  \right] \;\; \\ & B_z(x,y,z)  \approx
  \sum_{n=0}^\infty\sum_{m=0}^\infty \frac{x^n\,
  y^{2m+1}}{n!\,(2m+1)!}(-1)^m   \left[b^{[1]}_{n+2m} \right]
\end{split}\;\;,
\label{eq:linearfield}
\end{equation}
where the $n$ index has been shifted by 1 in the $B_x$ expansion for
convenience in the next step. Next, we arrange for $B_x(x,y,\Delta
L)=0$ by setting
\begin{equation}
  b^{[1]}_{n+2m} = -\frac{b^{[0]}_{n+2m}}{\Delta L} \;\;.
\label{eq:slope}
\end{equation}
It can be seen that this condition also assures $B_y(x,y,\Delta L)=0$.
This is a consequence of the requirement that $\nabla\times{\bf B}=0$.
Setting ${\bf B}(x,y,z)=0$ for $z\ge\Delta L$, we have assured that
the transverse field components are continuous.  Due to the artificial
assumption of linear fall-off of the field in the fringe region, the
longitudinal component $B_z$ is discontinuous in this approximation.

At this point, the ``multipole'' magnet has been idealized by a model
whose parameters, apart from its multipolarity index, are its
multipole strength $b^{[0]}_{n+2m}$, and its lengths $L_{\rm eff}$ and
$\Delta L_{\pm}$. This representation is appropriate for representing
the magnet within a particle tracking computer program. The lengths
$\Delta L_{\pm}$ could be determined by best-fitting to measured
fringe fields. But, to reduce the number of parameters in the
remainder of this paper, and with some reduction in accuracy, a
slightly different approach will be taken; the impulses delivered by
the fringe fields will be evaluated in a way that is independent of
the fringe field lengths: all the integrals involved will be computed
by using the ``hard-edge'' approximation, {\it i.e.} taking the limit
for which $\Delta L_\pm \rightarrow 0$. In this limit the straight
line approximation becomes exact.

For the sake of consistency another point must also be made. Since the
dominant multipole in the magnet body is also dominant in the fringe
field, there can be an appreciable contribution to the dominant field
integral (due to the magnet as a whole) that comes from the fields in
the fringe regions.  It is a matter of taste whether this contribution
is to be treated as part of the main field or part of the fringe
field.  In this paper, from here on, to simplify the formulas
somewhat, the term ``fringe field'' will refer to components other
than the dominant component, but restricted to those components
necessarily associated with the dominant multipole.   In other words,
the contributions from the dominant multipole component in the fringe
regions will be counted as part of the ideal magnet field
integral. Treating the magnet in this way increases its effective
length probably making it more nearly equal to the the physical magnet
length; {\it i.e.} $L \approx L_{\rm eff}$, and this will be assumed
in all subsequent formulas.

\section{DEFLECTIONS AT MAGNET ENDS}
\label{sec:scall}

For a given magnet with a perfect $2(n+1)$-pole geometry written in
cylindrical coordinates (see Appendix A), the scalar potential
satisfies the following symmetry condition:
\begin{equation}
\Phi(r,\theta,z)=\Phi(r,{\frac{\pi}{n+1}}-\theta,z) \;\;,
\label{eq:symrel}
\end{equation}
which leads to a relation between the harmonic multipole number
allowed by symmetry $n'$ and the multipole order $(n+1)$:
\begin{equation}
n'=(2j+1)(n+1)-1\;\;.
\label{eq:index}
\end{equation}
Thus, for a normal ``dipole'' ($n=0$) the multipole coefficients
allowed by the magnet symmetry are of the form $b_{2j}$, for a normal
``quadrupole'' ($n=1$) $b_{4j+1}$, for a normal ``sextupole'' ($n=2$)
$b_{6j+2}$, etc. Consider now a ``multipole magnet'', with normal
symmetry, for example. Following the symmetry
condition~(\ref{eq:index}), we can rewrite the field
components~(\ref{eq:cartcomp}), keeping terms of the expansion to
leading order:
\begin{equation}
\begin{split}
 B_x(x,y,z) =& {\mathcal Im}  \left\{  \frac{(x+iy)^n b_n(z) }{n!}
- {\frac{ (x+iy)^{n+1} \left[ (n+3)x - i(n+1)y\right]
b^{[2]}_{n}(z)}{4(n+2)!}} + O(n+4) \right\} \\  B_{y}(x,y,z) =&
{\mathcal Re} \left\{\frac{(x+iy)^n b_n(z)}{n!}
- {\frac{ (x+iy)^{n+1} \left[ (n+1)x - i(n+3)y\right]
b^{[2]}_{n}(z)}{4 (n+2)!}} +  O(n+4) \right\} \\ B_z(x,y,z) =&
{\mathcal Im} \left\{  {\frac{(x+iy)^{n+1} b^{[1]}_{n}(z)}{(n+1)!}}
+ O(n+3) \right\} \\
\end{split}
\;,
\label{eq:leadcomp}
\end{equation}
where the functions $O(j)$ represent polynomial terms in the
transverse variables $x,y$ of order greater or equal to $j$. These
expressions apply for $n>0$. The special case of the ``dipole'' will
be treated separately. Here the terms proportional to $b^{[1]}_{n}$
and $b^{[2]}_{n}$ approximate the fields present due to the
longitudinal field profile variation and do not include fields that
could be present due to non-ideal magnet design.

For a particle traversing the magnet along the straight line having
transverse coordinates $(x,y)$, the impulse ({\it i.e.} change of
transverse momentum) imparted by the nominal field component is
\begin{equation}
\begin{array}{lll}
\Delta p^{b}_{x}= -& e \int\limits_{\text{body}}   B_y(x,y,z) dz
\approx -& {\displaystyle { e   \overline{b_n} L_{\text{eff}}
{\frac{{\mathcal Re}\left\{ (x+iy)^n\right\}}{n!}} } }  \\ \Delta
p^{b}_{y}=  & e \int\limits_{\text{body}}   B_x(x,y,z) dz  \approx &
{\displaystyle { e    \overline{b_n} L_{\text{eff}}  {\frac{{\mathcal
Im}\left\{ (x+iy)^n\right\}}{n!}} }}
\end{array}\;\;,
\label{eq:intbody}
\end{equation}
where $ L_{\text{eff}}=\int\limits_{\text{body}} b_n(z)
dz/\overline{b_n}$ is the  effective length of the magnet, and
$\overline{b_n} $ is the nominal field coefficient in the body of the
multipole magnet.  The quantities in Eq.~(\ref{eq:intbody}), the
intentional and dominant (``zero order'') deflections caused by the
magnet, are only approximate, since they account neither for orbit
curvature within the body of the magnet nor for end field
deflections. Expressions like this will be used only as ``normalizing
denominators'' in ratios having (the presumably much smaller) magnet
end deflections as numerators.  For magnets other than bending
magnets, for which the average deflection is zero, it will be
necessary to use r.m.s. values for both the normalizing denominator
and the numerator.

The impulse due to the fringe field at one end of a magnet is defined
in this paper as the effect of field deviation from nominal, from well
inside (where the nominal multipole coefficient is assumed to be
independent of $z$) to well outside the magnet (where all field
components are assumed to vanish). These will be the limits for the
integrals used in order to calculate the fringe deflection.  To obtain
explicit formulas the upper limit of these integrals will be taken to
be infinity. Exploiting the assumed constancy of $x$ and $y$ along the
orbit, these integrals will all be evaluated using integration by
parts.

Suppressing the entire pure multipole contribution, as explained
above, we have $\int_{-\infty}^{\infty}{\bf B}(x,y,z)dz \approx 0$.
For $x=y=0$ this is an equality {\it by definition\/}, and for finite
displacements it is approximately true if, as we are assuming, the
transverse particle displacements remain approximately constant.  This
is consistent with our straight line orbit approximation.

The individual components of the impulse can themselves be separated
into terms due to longitudinal fields (labeled $\parallel$) and due to
transverse fields (labeled $\perp$);
\begin{equation}
\Delta p^f_{x,y} = \Delta p^f_{x,y}(\parallel) + \Delta
p^f_{x,y}(\perp) \;\;,
\end{equation}
where
\begin{equation}
\begin{array}{ll}
\Delta p^f_{x}(\parallel) =  & e \int\limits_{\text{fringe}}   y'
B_z(x,y,z) dz \\ \Delta p^f_{y}(\parallel) = -& e
\int\limits_{\text{fringe}}   x' B_z(x,y,z) dz
\end{array}
\label{eq:longcomp}
\end{equation}
are the momentum increments of the particle caused by the longitudinal
component of the magnetic field and
\begin{equation}
\begin{array}{ll}
\Delta p^f_{x}(\perp) =  -& e \int\limits_{\text{fringe}}   B_y(x,y,z)
dz \\ \Delta p^f_{y}(\perp) =   & e \int\limits_{\text{fringe}}
B_x(x,y,z) dz
\end{array}
\label{eq:transcomp}
\end{equation}
are the momentum increments of the particle caused by the transverse
components of the magnetic field.  Using the leading order expressions
of the magnetic field, we obtain the relations
\begin{equation}
\begin{array}{ll}
\Delta p^f_{x}(\parallel) \approx   &  {\displaystyle {\frac{e
  \overline{b_n}}{(n+1)!}  {\mathcal Im}\left\{
  (x+iy)^{n+1}\right\}}y'  }\\  & \\ \Delta p^f_{y}(\parallel) \approx
  - &  {\displaystyle {\frac{e \overline{b_n}}{(n+1)!}  {\mathcal
  Im}\left\{ (x+iy)^{n+1}\right\}}x' }
\end{array}
\label{eq:multlong}
\;\;,
\end{equation}
and
\begin{equation}
\begin{array}{ll}
\Delta p^f_{x}(\perp) & \approx  {\displaystyle {\frac{-e
\overline{b_n}}{4(n+1)!}}  {\mathcal Re}\left\{ (x+iy)^{n}
\left[(n+1)xx'+(n+3)yy'+i(n-1)xy'-i(n+1)yx'\right]\right\} }\\ &  \\
\Delta p^f_{y}(\perp) & \approx  {\displaystyle{\frac{e
\overline{b_n}}{4(n+1)!}}  {\mathcal Im}\left\{ (x+iy)^{n}
\left[(n+3)xx'+(n+1)yy'+i(n+1)xy'-i(n-1)yx'\right]\right\} }
\end{array}
\label{eq:multtrans}
\;.
\end{equation}

The total impulses caused by the fringe field are therefore
\begin{equation}
\begin{array}{lll}
\Delta p^f_{x} & \approx - & {\displaystyle  {\frac{e
\overline{b_n}}{4(n+1)!}}  {\mathcal Re}\left\{ (x+iy)^n
\left[(n+1)(x-iy)(x'+iy')+2iy'(x+iy)\right]\right\} }\\ & & \\ \Delta
p^f_{y} & \approx   & {\displaystyle{\frac{e
\overline{b_n}}{4(n+1)!}}  {\mathcal Im}\left\{ (x+iy)^n
\left[(n+1)(x-iy)(x'+iy')-2x'(x+iy)\right]\right\} }
\end{array}
\label{eq:fringmult}
\;\;.
\end{equation}

Even though they occur at a fixed point in the lattice, because these
impulses depend on slopes $x'$ and $y'$ and are truncated Taylor
series, they are not symplectic. To use them in long term,
damping-free tracking, symplecticity would have to be restored by
including deviations in transverse
coordinates~\cite{ForestMilut,Irwin,Forest,Baartman}.

\section{APPLICATION EXAMPLES}

The formulas just derived are appropriate to calculate the end field
deflection  of any single particle. But to assess the importance of
these deflections it is appropriate to calculate their impact on the
beam as a whole, for example by calculating an r.m.s. deflection, such
as $(\Delta p^f_{\perp})_{rms}= \sqrt{\langle(\Delta p^f_x)^2\rangle +
\langle(\Delta p^f_y)^2\rangle}$.  Here the operator $\langle
. \rangle$ denotes an averaging over angle variables. Note that here,
and from here on, the subscript $\perp$ specifies the transverse
impulse, and does not  refer to a magnetic field component.  Formulas
for r.m.s. values like these are derived in Appendix~B.  This section
contains examples of the use of those formulas, starting with the
cases of flat and round beams, then specializing the results further
for ``dipole'' and ``quadrupole'' magnets. The derived formulas are
finally applied for evaluating the impact of magnets end fields in the
case of the Large Hadron Collider (LHC) and the Spallation Neutron
Source (SNS) accumulator ring. The calculations are based on
Eq.~\eqref{eq:finkick}.

\subsection{Flat Beam }

For a flat beam, one of the transverse degrees of freedom ({\it e.g.}
the vertical $y,y'$) vanishes. Thus, the total transverse
r.m.s. momentum increment from the magnet body is
\begin{equation}
(\Delta p^b_{\perp})_{\text rms}\equiv \sqrt{\langle(\Delta
 p^b_x)^2\rangle} \approx \frac{e   \overline{b_n}
 L_{\text{eff}}}{2^{n}n!}\sqrt{\binom{2n}{n}
 \overline{\beta^n}\epsilon_{\perp}^n}
\label{eq:rmsbodflat}
\;\;,
\end{equation}
where $\overline{\beta^n}$ represents the average of the $\beta^n$ in
the body of the magnet and $\epsilon_{\perp}$ is the transverse
emittance. The total transverse r.m.s. momentum increment from one of
the fringes of the magnet is
\begin{equation}
(\Delta p^f_{\perp})_{\text rms}\equiv \sqrt{\langle(\Delta
 p^f_x)^2\rangle} \approx \frac{e  \overline{b_n}}{2^{n+3}n!}
 \sqrt{\binom{2n+2}{n+1}\frac{\beta^n[1+(2n+3)\alpha^2]}{2(n+2)}
 \epsilon_{\perp}^{n+2}}
\label{eq:rmsfrinflat}
\;\;,
\end{equation}
where $\beta$ and $\alpha$ represent the beta and alpha functions, at
the fringe location. The ratio of these quantities is
\begin{equation}
\frac{(\Delta p^f_{\perp})_{\text rms}}{(\Delta p^b_{\perp})_{\text
 rms}} \approx \frac{\epsilon_{\perp}}{8 L_{\text{eff}}}
 \sqrt{\frac{(2n+1)\beta^n[1+(2n+3)\alpha^2]}{(n+1)(n+2)\overline{\beta^n}}}
 \;\;.
\label{eq:ratrmsflat}
\end{equation}
Assuming that the beta functions are not varying rapidly, if the
magnets are in non-critical locations (which is to say most magnets),
the square root dependence can be neglected, so an order-of-magnitude
estimate (dropping an $n$-dependent numerical factor not very
different from 1) is given by
\begin{equation}
\frac{(\Delta p^f_{\perp})_{\text rms}}{(\Delta p^b_{\perp})_{\text
 rms}}  \approx \frac{\epsilon_{\perp}}{L_{\text{eff}}} \;\;.
\label{eq:ratrmsflatemlen}
\end{equation}

The case in which fringe field deflections are likely to be most
important is when $\alpha$ is anomalously large, for example in the
vicinity of beam waists such as at the location of intersection points
in colliding beam lattices. In this case, (again dropping a numerical
factor) the ratio of deflections is roughly
\begin{equation}
\frac{(\Delta p^f_{\perp})_{\text rms}}{(\Delta p^b_{\perp})_{\text
 rms}}  \approx \alpha \frac{ \epsilon_{\perp}}{L_{\text{eff}}} \;\;.
\label{eq:ratrmsflatalemlen}
\end{equation}
The same result is obtained by setting $\beta_x>>\beta_y$ in
Eqs.~(\ref{eq:finkick}).

Often the relative deflection is so small as to make neglect of  the
fringe field deflection entirely persuasive. The simplicity of the
formula  is due to the fact that the fringe contribution is expressed
as a fraction of the dominant contribution.  Note that, as stated
before, this formula applies to each end separately, and does not
depend on any cancellation of the contributions from two ends. In
fact, nonlinear analysis shows that in magnets fringe-field
contributions can tend to add up instead of cancelling~\cite{Forest}.

\subsection{Round beam}

\begin{figure}
\center
\mbox{\epsfig{file=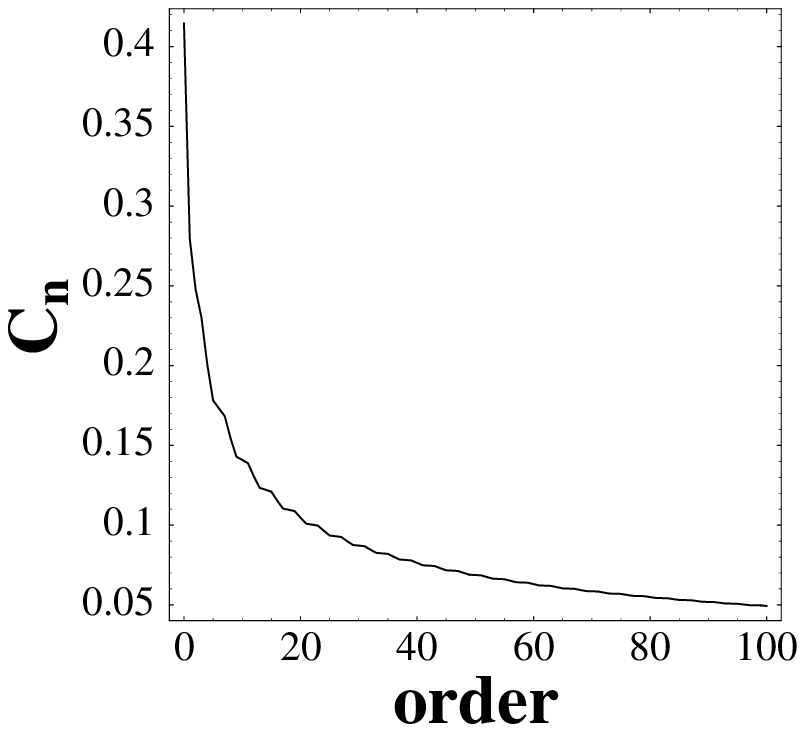,width=0.4\columnwidth} \hskip 1pt
\epsfig{file=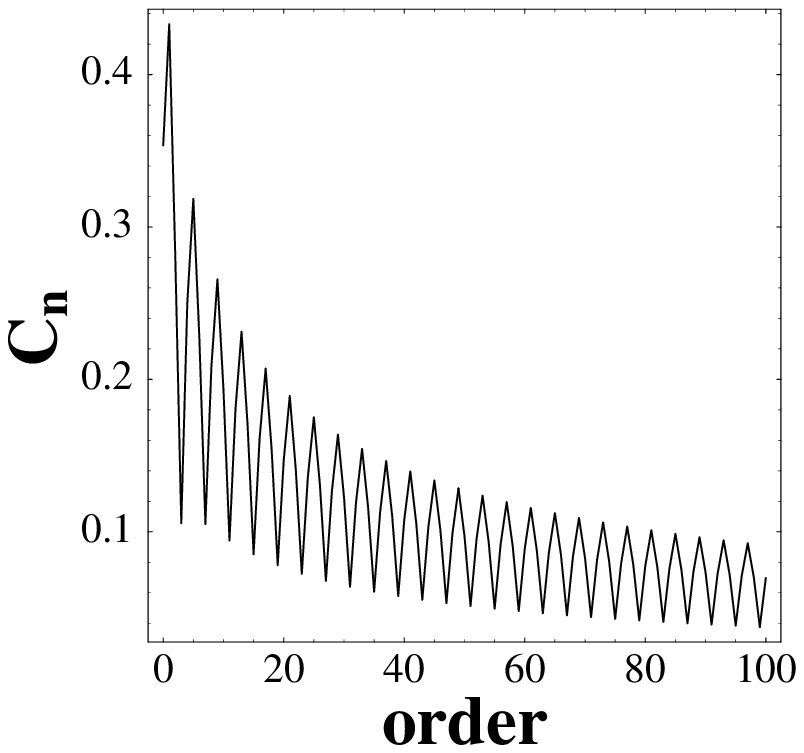,width=0.4\columnwidth}} 
\vskip .2cm
\caption{Order dependent coefficient of the momentum increment ratio,
for a round beam (see Eq.~(\ref{eq:cnround})), when the $\alpha$
function is small (left) and when the $\alpha$ function is large
(right).}
\label{fig:roundcoeff}
\end{figure}

For a round beam, the two transverse emittances are equal
$\epsilon_x=\epsilon_y=\epsilon_{\perp}$.  For simplicity, we assume
that typical values of horizontal and vertical lattice functions are
approximately equal; $\beta_x \approx \beta_y = \beta$ and $\alpha_x
\approx \alpha_y = \alpha$. Also assume that $\overline{\beta^n}
\approx \overline{\beta}^n$, {\it i.e.} the beta functions do not vary
significantly in the body of the magnet. Taking into account the
previous hypotheses,  the total transverse r.m.s.  momentum increment
for the body becomes:
\begin{equation}
(\Delta p^b_{\perp})_{\text rms} \approx \frac{e   \overline{b_n}
 L_{\text{eff}}}{2^{n/2}n!}
 \overline{\beta}^{n/2}\epsilon_{\perp}^{n/2}
 \left[_3F_2(1/2,-n,-n;1,1/2-n;1)\frac{(2n-1)!!}{n!}\right]^{1/2}
\label{eq:rmsbodround}
\;\;,
\end{equation}
where the function in the square root represents the generalized
Hyper-geometric function  (see~\cite{Gradshteyn} for
details). Applying the same simplifications, the r.m.s. momentum kick
given by the fringe field is:
\begin{equation}
(\Delta p^f_{\perp})_{\text rms}  \approx \frac{e
\overline{b_n}\beta^{n/2}\epsilon_{\perp}^{n/2+1}}{2^{n+3}(n+1)!}
\left[ \sum_{l=0}^{n} \binom{2(n-l)}{n-l} \binom{2l}{l}
g_{n,l}(\alpha^2)   \right]^{1/2}
\label{eq:rmsfrinround}
\;\;,
\end{equation}
where we considered $\beta_x \approx \beta_y = \beta$ and the same for
the $\alpha$ functions. Notice now that the sum of the coefficients
$g_{n,l}=g_{n,l,0}+g_{n,l,1}+g_{n,l,2}$ depends only on
$\alpha^2$. The series involving them can be also written as a sum of
a few generalized Hyper-geometric functions. The ratio of the
r.m.s. momentum transverse kicks is:
\begin{equation}
\frac{(\Delta p^f_{\perp})_{\text rms}}{(\Delta p^b_{\perp})_{\text
 rms}} \approx \frac{\epsilon_{\perp}}{L_{\text{eff}}}
 \frac{\beta^{n/2}}{\overline{\beta}^{n/2}} C_n(\alpha^2) \;\;,
\label{eq:ratrmsround}
\end{equation}
where the coefficient $C_n$ is:
\begin{equation}
C_n(\alpha^2) = \frac{1}{8(n+1)} \left[ \frac{n! \sum_{l=0}^{n}
\binom{2(n-l)}{n-l} \binom{2l}{l}
g_{n,l}(\alpha^2)}{_3F_2(1/2,-n,-n;1,1/2-n;1)(2n-1)!!} \right]^{1/2}
\;\;.
\label{eq:cnround}
\end{equation}
Let us consider two cases, as before: one where $\alpha$ is small and
one where $\alpha$ is large, as near the interaction points of large
colliders. For the first case ($\alpha$ small), we may neglect the
terms having $\alpha$ as a factor in the coefficient $g_{n,l}$ and in
the second case, we can pull out $\alpha$ from the square root and
neglect terms in the coefficient $g_{n,l}$ having now the $\alpha$
function in the denominator. In this way, the coefficients $C_n$ of
Eq.~\eqref{eq:cnround} will depend only on the order $n$. We plot in
Figs.~\ref{fig:roundcoeff}, the behavior of these coefficients as a
function of the multipole order $n$, for large and small $\alpha$. The
dominant factor in $C_n$ seems to be $1/(n+1)$, which is reflected in
the slow asymptotic decay depicted at the plots. For all practical
cases (multipole orders up to 20),  $C_n$ lies between 1/2 and
1/10. Assuming now that the average $\beta$ in the body of the magnet
is not so different from $\beta$ in the fringe, one gets for small
$\alpha$ functions:
\begin{equation}
\frac{(\Delta p^f_{\perp})_{\text rms}}{(\Delta p^b_{\perp})_{\text
 rms}}  \approx  \frac{ \epsilon_{\perp}}{L_{\text{eff}}} \;\;,
\label{eq:ratrmsbroundemlen}
\end{equation}
as in Eq.~(\ref{eq:ratrmsflatemlen}), and for $\alpha$ large:
\begin{equation}
\frac{(\Delta p^f_{\perp})_{\text rms}}{(\Delta p^b_{\perp})_{\text
 rms}} \approx  \alpha \frac{\epsilon_{\perp}}{L_{\text{eff}}} \;\;,
\label{eq:ratrmsbroundalemlen}
\end{equation}
as in Eq.~(\ref{eq:ratrmsflatalemlen}).

\subsection{Dipole magnet}

Consider a ``straight'' dipole magnet; the configuration of poles and
coils is symmetric about the $x=0$ and $y=0$ planes, and the coils are
excited with alternating signs and equal strength.  By symmetry $B_x$
is odd in both $x$ and $y$, $B_y$ is even in  both $x$ and $y$, and
$B_z$ is even in $x$ and odd in $y$.  Using the general field
expansion of Eq.~(\ref{eq:fieldmgen}), we get:
\begin{equation}
\begin{split}
B_x & =  {\displaystyle \sum^{\infty}_{m,n=0}\sum^{m}_{l=0}
{\frac{(-1)^m x^{2n+1}y^{2m+1}}{(2n+1)!(2m+1)!}}  \binom{m}{l}
b^{[2l]}_{2n+2m+2-2l} }\\  B_y & = {\displaystyle
\sum^{\infty}_{m,n=0}\sum^{m}_{l=0} {\frac{(-1)^m
x^{2n}y^{2m}}{(2n)!(2m)!}}  \binom{m}{l} b^{[2l]}_{2n+2m-2l} }\\  B_z
& =  {\displaystyle  \sum^{\infty}_{m,n=0}\sum^{m}_{l=0} {\frac{(-1)^m
x^{2n}y^{2m+1}}{(2n)!(2m+1)!}} \binom{m}{l} b^{[2l+1]}_{2n+2m-2l} }\\
\end{split}\;\;.
\label{eq:fdipole}
\end{equation}
Taking the field expansion up to leading order, we get:
\begin{equation}
\begin{array}{lll}
B_x \!\! & \!\!\;=\;\!\! &\!\! {\displaystyle b_{2} xy  + O(4)}\\ B_y
\!\!& \!\!\;=\;\!\! & \!\!{\displaystyle  b_{0} - \frac{1}{2}
b_{0}^{[2]} y^2 + \frac{1}{2} b_2 (x^2 - y^2) + O(4)}\\ B_z \!\!&
\!\!\;=\;\!\! & \!\!{\displaystyle y \; b^{[1]}_{0} \!+\! O(3)}\\
\end{array}
\;\;,
\label{eq:dexpand2}
\end{equation}
where $b_2$ represents a sextupole field component allowed by the
symmetry of the ``dipole'' magnet (for an ideally designed magnet
$b_2=0$) and $O(3)\;\text{and}\;O(4)$ contain all the allowed terms of
higher orders.

A point has to be made about the application of the integrals
evaluating the rms momentum kicks for bending magnets: because of the
curved central orbit, these integrals are not exact, as previously
mentioned. Nevertheless, in most practical cases, the field uniformity
in the interior of a ``dipole'' magnet is very high, and thus, on
heuristic grounds, this approach can be expected to provide fairly
good estimates even in this case.

The change of transverse momentum imparted by the dipole field is (see
Eq.~(\ref{eq:intbody}))
\begin{equation}
\Delta p^b= -e \int\limits_{\text{body}}   b_0 dz \approx -e
\overline{b_0} L_{\text{eff}} \;\;,
\end{equation}
where as before $L_{\text{eff}}=\int_{\text{body}} b_0
dz/\overline{b_0}$ is the effective  length of the ``dipole'' magnet,
and $\overline{b_0}$ is the main dipole field in the body of the
``dipole'' magnet. Using Eq.~(\ref{eq:transcomp}) the deflections in
one fringe are
\begin{equation}
\Delta p^f_{x} \approx  2 e   \overline{b_0} y y' \;\;,\qquad \Delta
p^f_{y} \approx  - e   \overline{b_0} y x' \;\;,
\end{equation}
and the total r.m.s. fringe kick is
\begin{equation}
(\Delta p^f_{\perp})_{\text rms} = e   \overline{b_0} \sqrt{ 4 \langle
y^2 {y'}^2\rangle +\langle y^2 {x'}^2\rangle}\;\;.
\end{equation}
Using Eqs.~(\ref{eq:emitang}) and~(\ref{eq:averaging}), we have
\begin{equation}
\langle y^2 {y'}^2\rangle  = \frac{(1+3\alpha_y^2)\epsilon_{y}^2}{8}
\;\;,\qquad  \langle y^2 {x'}^2\rangle  = \langle y^2 \rangle
\langle{x'}^2\rangle   =
{\frac{(1+\alpha_x^2)\beta_y\epsilon_{x}\epsilon_{y} }{4\beta_x}} \;\;,
\end{equation}
and the r.m.s. transverse momentum kick becomes
\begin{equation}
(\Delta p^f_{\perp})_{\text rms} = e   \overline{b_0}
\sqrt{\frac{(1+3\alpha_y^2)\epsilon_{y}^2}{8}+
\frac{(1+\alpha_x^2)\beta_y\epsilon_{x}\epsilon_{y} }{4\beta_x}} \;\;,
\end{equation}
Thus, the by-now-standard ratio is
\begin{equation}
\frac{(\Delta p^f_{\perp})_{\text rms}}{(\Delta p^b_{\perp})_{\text
 rms}} \approx {\displaystyle
 \frac{1}{L_{\text{eff}}}}\sqrt{\frac{(1+3\alpha_y^2)\epsilon_{y}^2}{8}+
 \frac{(1+\alpha_x^2)\beta_y\epsilon_{x}\epsilon_{y}}{4\beta_x}} \;\; .
\label{dipdifl}
\end{equation}
Except for numerical factors near one this formula yields the same
``ball-park'' estimates as given by Eq.~(\ref{eq:ratrmsbroundemlen})
and Eq.~(\ref{eq:ratrmsbroundalemlen}) for the small $\alpha$ and
large  $\alpha$ cases.

\subsection{``Quadrupole'' magnet}

The configuration of poles and coils in a ``quadrupole'' magnet is
symmetric about the four planes $x=0;\ y=0;\ x=y;\ x=-y$ and if the
coils are excited with alternating signs and equal strength, the
magnetic field will satisfy the following symmetry conditions: $B_x$
is even in $x$ and odd in $y$; $B_y$ is odd in $x$ and even in $y$;
$B_z$ is odd in both $x$ and $y$; and $B_z(x,y,z)=B_z(y,x,z)$.  As
before, we may express the field components as:
\begin{equation}
\begin{split}
B_x  = & {\displaystyle \sum^{\infty}_{m,n=0}\sum^{m}_{l=0}
{\frac{(-1)^mx^{2n}y^{2m+1}}{(2n)!(2m+1)!}}  \binom{m}{l}
b^{[2l]}_{2n+2m+1-2l} }\\  B_y = & {\displaystyle
\sum^{\infty}_{m,n=0}\sum^{m}_{l=0}
{\frac{(-1)^mx^{2n+1}y^{2m}}{(2n+1)!(2m)!}}  \binom{m}{l}
b^{[2l]}_{2n+2m+1-2l}  }\\ B_z = & {\displaystyle
\sum^{\infty}_{m,n=0}\sum^{m}_{l=0}
{\frac{(-1)^mx^{2n+1}y^{2m+1}}{(2n+1)!(2m+1)!}} \binom{m}{l}
b^{[2l+1]}_{2n+2m+1-2l}  }\\
\end{split}.
\label{eq:expand1}
\end{equation}
The field expansion can be written as
\begin{equation}
\begin{array}{lll}
B_x & = & {\displaystyle y \left[b_1-\frac{1}{12}(3x^2+y^2)b_1^{[2]}
\right] + O(5)}\vspace{.1cm} \\  B_y & = & {\displaystyle x
\left[b_1-\frac{1}{12}(3y^2+x^2)b_1^{[2]} \right] + O(5)}\\ B_z & = &
{\displaystyle xy b_1^{[1]} + O(4)}\\
\end{array}
\;\;,
\label{eq:expand2}
\end{equation}
where $ b_1(z)$ is the transverse field gradient at the quadrupole
axis, and $O(4),O(5)$ contain all the higher order terms. For a
particle traversing the magnet with a horizontal deviation $x$ and
vertical deviation $y$ from the center, the momentum increments
produced by the nominal field gradients are
\begin{equation}
\Delta p^b_{x}=  -e   \overline{b_1}  x L_{\text{eff}}\;\;, \qquad
\Delta p^b_{y}=   e   \overline{b_1}  y L_{\text{eff}} \;\;,
\end{equation}
where $L_{\text{eff}}=\int_{\text{body}} b_1 dz/\overline{b_1}$ is the
effective  length of the quadrupole magnet.  The momentum increments
of the particle contributed from the longitudinal component of the
magnetic field are
\begin{equation}
\Delta p^f_{x}(\parallel) \approx   e   x y y' \overline{b_1} \;\;,
\qquad \Delta p^f_{y}(\parallel) \approx - e   x y x'\overline{b_1}
\;\;,
\end{equation}
and the momentum increment produced by the transverse component of the
fringe fields are
\begin{equation}
\Delta p^f_{x}(\perp) \approx \frac{-e   \overline{b_1}}{4} \left[ 2 x
y y' + (x^2+y^2)x' \right] \;\;, \qquad \Delta p^f_{y}(\perp) \approx
\frac{e   \overline{b_1}}{4}  \left[ 2 x x'y  + (x^2+y^2)y'\right]
\;\;.
\end{equation}
Combining the contributions, the total momentum increments due to
fringe field are
\begin{equation}
\begin{split}
\Delta p^f_{x}\approx & \frac{e   \overline{b_1}}{4} \left[ 2 x y y' -
(x^2+y^2)x' \right]\\ \Delta p^f_{y} \approx & \frac{e
\overline{b_1}}{4}  \left[-2 x x'y +  (x^2+y^2)y'\right]
\end{split}\;\;.
\end{equation}
Again, by averaging the sum of squares of the transverse momenta
contribution,  we obtain the total rms transverse momentum kick
imparted by the fringe field:
\begin{equation}
\begin{array}{ll}
{\displaystyle (\Delta p^f_{\perp})_{\text rms} \approx \frac{e
\overline{b_1}}{16}    } & {\displaystyle \left\{\;
(1+5\alpha_x^2)\beta_x\epsilon_x^3 + \frac{3}{\beta_y} \left[
(1+\alpha_y^2)\beta_x^2-8\alpha_x\alpha_y\beta_x\beta_y+2(1+3\alpha_x^2)\beta_y^2
\right] \epsilon_x^2 \epsilon_y \right.} \\ & {\displaystyle \left.  +
(1+5\alpha_y^2)\beta_y\epsilon_y^3 + \frac{3}{\beta_x} \left [
(1+\alpha_x^2)\beta_y^2-8\alpha_x\alpha_y\beta_x\beta_y
+2(1+3\alpha_y^2)\beta_x^2   \right] \epsilon_x \epsilon_y^2
\right\}^{1/2} }
\end{array}.
\end{equation}
Note that the expected rotation symmetry of the quadrupole is
exhibited both in this formula and in the body deflection formula.
The standard ratio is
\begin{equation}
\begin{array}{l}
{\displaystyle \frac{(\Delta p^f_{\perp})_{\text rms}}{(\Delta
p^b_{\perp})_{\text rms}}  \approx \frac{1}{8{L_{\text{eff}}}}}
{\displaystyle \left\{ \frac{
(1+5\alpha_x^2)\beta_x^2\beta_y\epsilon_x^3 + 3\beta_x \left[
(1+\alpha_y^2)\beta_x^2-8\alpha_x\alpha_y\beta_x\beta_y+2(1+3\alpha_x^2)\beta_y^2
\right] \epsilon_x^2 \epsilon_y}{2\beta_x\beta_y
(\overline{\beta_x}\epsilon_x +\overline{\beta_y}\epsilon_y)} \right.}
\\ {\displaystyle \left.  +
\frac{(1+5\alpha_y^2)\beta_x\beta_y^2\epsilon_y^3 + 3\beta_y \left[
(1+\alpha_x^2)\beta_y^2-8\alpha_x\alpha_y\beta_x\beta_y
+2(1+3\alpha_y^2)\beta_x^2   \right] \epsilon_x
\epsilon_y^2}{2\beta_x\beta_y (\overline{\beta_x}\epsilon_x
+\overline{\beta_y}\epsilon_y)} \right\}^{1/2} }
\end{array}
\;\;.
\label{quaddifl}
\end{equation}
Again dropping factors near 1, this leads to the same ball-park
estimates of Eq.~(\ref{eq:ratrmsbroundemlen})  and
Eq.~(\ref{eq:ratrmsbroundalemlen}) .

\subsection{Magnets of LHC and SNS}

The LHC and the SNS accumulator ring are good examples for testing the
validity of the derived fringe field figure of merit formulas. Indeed,
the purpose of these two proton machines and thereby their magnet
design differs in great extent: the LHC, a high-energy hadron
collider, is filled with long super-conducting magnets of very small
aperture (around 1 cm). In contrast, the SNS ring, a low-energy high
intensity accumulator, contains short normal conducting magnets with
wide aperture (tens of cm). In addition, the lattice design, optics
functions and physical parameters of the two machines are
substantially different, e.g. the emittance of the SNS beam is several
orders of magnitude bigger, than the one of the LHC. In
Table~\ref{tab:param}, we summarize the parameters of the main magnets
in the two accelerators entering in the figure of merit formulas
~\eqref{dipdifl} and~\eqref{quaddifl}.

\begin{figure}
\center
\epsfig{file=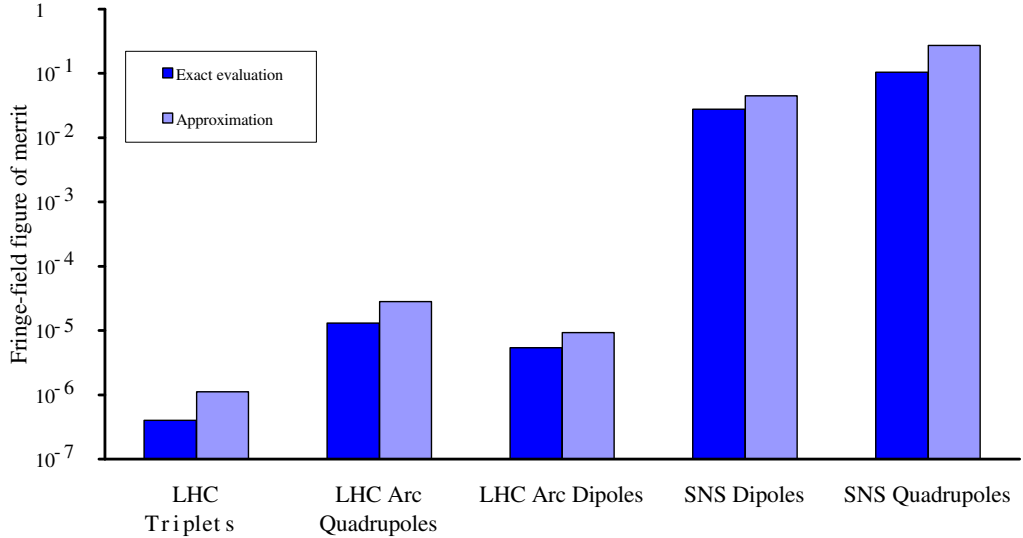,width=1\columnwidth}
\caption{Fringe-field figure of merit estimates for the LHC and the
SNS accumulator ring magnets.  The two different bars in each case
represent evaluation with the exact formulas derived for the
fringe-field figure of merit of Eqs.~\eqref{dipdifl}
and~\eqref{quaddifl} (black bars) and the  approximate formula for
round beams~\eqref{eq:ratrmsbroundemlen} (grey bars).}
\label{fig:fringeLHC}
\end{figure}

In Fig.~\ref{fig:fringeLHC}, we plot in logarithmic scale the
fringe-field figure of merit estimates for the LHC and the SNS
accumulator ring magnets.  The black bars  represent evaluation with
the exact formulas derived for dipoles and quadrupoles (see
Eqs.~\eqref{dipdifl} and~\eqref{quaddifl}) and the grey bars represent
the evaluation with the formula for round
beams~\eqref{eq:ratrmsbroundemlen}. In both cases, the total effect
for each magnet is computed by summing up the fringe-field figures of
merit from both ends due to all the magnets of the same type. The
fringe field importance in the case of the SNS is striking, especially
for quadrupole magnets, whereas in the case of the LHC can be
completely neglected. Note that similar results can be derived by
careful dynamical analysis and computation of tune-shifts due to
fringe fields or dynamic aperture analysis for both the
LHC~\cite{Meot} and the SNS~\cite{Papaphilippou}.  It is important to
stress that even the approximate formula for round
beams~\eqref{eq:ratrmsbroundemlen} is slightly pessimisitic and within
a factor of 2 of the exact figure of merit.

\section{Conclusion}

We have derived formulas for the momentum kicks imparted by the fringe
fields of general straight (non-solenoidal) multipole magnets. These
formulas are based on an expansion having arbitrary dependence on the
longitudinal coordinate.  This expansion can be used for direct
integration of the equations of motion for particle tracking  or other
analytical non-linear dynamics estimates. It also permits the fringe
part  and the body part of individual magnets to be identified and
separated.  A figure of merit, the ratio of r.m.s. end deflection to
r.m.s. body deflection is introduced and evaluated. Its
proportionality to the  transverse emittance results in an
easily-evaluated measure of the importance of fringe fields both in
cases in which the variation of optical functions is not too rapid and
in the opposite case of rapid variation.  These results are in
agreement with previous crude estimations which employed simple
physics arguments based on Maxwell laws~\cite{theorem}. Finally, the
formalism has been applied to the most common cases of multipole
magnets, namely normal ``dipoles'' and
``quadrupoles''~\cite{EPAC2000}. Since the straight line approximation
has been used throughout, these formulas are only precise for magnetic
fields that are well-approximated by step functions (the ``hard-edge''
approximation).  Thus, the formulas contain no parameters associated
with the fringe shape (for example,
see~\cite{Venturini,BerzErdelyiMakino}).  Also, as stated previously,
only those fringe fields matching, and therefore required by, the
nominal body multipolarity are accounted for.

Numerical evaluation of the end/body figure of merit shows that fringe
fields can be neglected in the magnets populating the arcs of large
colliders like the LHC. In these rings, the magnets are long enough
and the emittances are so small (of the order of $10^{-9}$~m~rad) that
the effect of fringe fields is a tiny perturbation as compared to the
dominant multipole errors in the body of the magnets. The effect may
be important, however,  in small rings, as the SNS accumulator
ring~\cite{Papaphilippou} or the muon collider
ring~\cite{BerzErdelyiMakinoII}, where the emittance is large
(typically $10^{-4}$~m~rad) and the magnets much shorter. Careful
consideration should be also taken in the case of the magnets located
in the interaction regions of the collider~\cite{Wan}, where the beta
variation is quite big.

It is perhaps appropriate to call attention to possible
``overly optimistic'' use of the scaling law. Often quadrupoles
are grouped in doublets or triplets in which the desired focal
properties rely on the intentional, highly-tuned, near cancellation
of deflections caused by more than one element. In such cases,
the fringe deflections are, of course, amplified, when evaluated
relative to the gross multiplet deflection. This effect
is most obvious at focal points. 

Since the early analytical studies of Lee-Whiting~\cite{LeeWhiting}
and Forest~\cite{ForestMilut,Forest}, significant progress has been
achieved for the construction of accurate maps which represent the
motion of particles through the magnet fringe field, using either
direct numerical evaluation with exact integration of the magnetic
field~\cite{VenturiniDragt,marylie} or parameter fit of an adequate
function~\cite{Hoffstater,ErderlyiLindermann,COSY} ({\it e.g.} the
Enge function~\cite{Enge}). These maps are essential for the study of
non-linearities introduced by fringe-fields through Hamiltonian
perturbation theory techniques. On the other hand, the scaling law we
have emphasized can provide a rough estimate of the impact of these
fringe fields in a ring. If the fringe fields are found to be
important, a thorough numerical modelling and analysis of their effect
has to be undertaken, including computation of the amplitude dependent
tune-shift, resonance excitation and dynamic
aperture~\cite{Venturini,BerzErdelyiMakino,Papaphilippou,PEPII,Zimmermann,ZimmermannII,ZimmermannIII,BerzErdelyiMakinoIII},
as non-linear dynamics can be very sensitive to the details of
different lattices and magnet designs. Furthermore, great care is
required to preserve symplecticity and use these maps in particle
tracking.

\appendix

\section{3D MULTIPOLE EXPANSION, CYLINDRICAL COORDINATES}

The magnetic field representation in Cartesian coordinates $(x,y,z)$
is not optimal for studying symmetries imposed by the cylindrical
geometry of a perfect multipole magnet. For this, it is preferable to
rely on expansions in cylindrical coordinates
$(r,\theta,z)=(\sqrt{x^2+y^2},\arctan{(y/x)},z)$
\cite{Danby,Brown,Bassetti,ForestMilut,Forest,Gardner,VenturiniDragt}.
Both expansions are equivalent and the use of the former or the latter
depends mostly on taste and the specific problem to be treated.

First, consider the magnetic scalar potential written in the following
form~\cite{ForestMilut,Forest}
\begin{equation}
  \Phi(r,\theta,z)= {\mathcal Re} \left \{ \sum_{n=0}^\infty e^{i
(n+1)\theta} \sum_{m=0}^\infty {\cal G}_{n+1,m}(z) r^m \right \} \;\;,
\label{eq:potcyn}
\end{equation}
where now the $z$--dependent coefficients ${\cal G}_{n+1,m}(z)$ are
generally complex.  The above expansion follows directly from the fact
that the Laplacian commutes with
$\partial/\partial\theta$~\cite{Forest}. This allows the consideration
of solutions where the dependence in $\theta$ is an harmonic
$2(n+1)$-pole. This expansion is compatible with the general solution
of the Laplace equation in cylindrical coordinates, involving Bessel
functions~\cite{VenturiniDragt,Jackson,Jain}.

Using Eq.~(\ref{eq:potcyn}) and the Laplace equation, one gets that
${\cal G}_{n+1,0} = 0$. Moreover, ${\cal G}_{n+1,1}$, should vanish
for $n>0$ (all terms except the dipole). Finally, we have a recursion
relation~\cite{Forest,Gardner} similar to Eq.~(\ref{eq:coef}):
\begin{equation}
{\cal G}_{n+1,m+2}(z) = {\frac{{\cal
G}^{[2]}_{n+1,m}(z)}{(n+1)^2-(m+2)^2}} \quad \text{for} \quad m \ne
n-1 \;\; ,
\label{eq:cocyn}
\end{equation}
where again the superscript in brackets denotes derivatives with
respect to $z$. Following these relations, one can show that all
coefficients with $m<n+1$ vanish.  Thus, the first  non-zero
coefficient is ${\cal G}_{n+1,n+1}$ (for $m=n+1$). By extending the
recursion relation (\ref{eq:cocyn}) so as to express any coefficient
as a function of ${\cal G}_{n+1,n+1}$, we get:
\begin{equation}
{\cal G}_{n+1,n+1+2k}(z) = {\frac{(-1)^k (n+1)!}{2^{2k}(n+1+k)!k!}}
{\cal G}^{[2k]}_{n+1,n+1}(z)\;\; .
\label{eq:recu}
\end{equation}
The summation indexes can be rearranged so as to express the magnetic
scalar potential in cylindrical coordinates~\cite{Forest,Kareh}:
\begin{equation}
\Phi(r,\theta,z)= {\mathcal Re} \left \{ \sum_{n=0}^\infty e^{i
(n+1)\theta} \sum_{k=0}^\infty {\frac{(-1)^k
(n+1)!}{2^{2k}(n+1+k)!k!}} {\cal G}^{[2k]}_{n+1}(z) \, r^{n+1+2k}
\right \} \;\;,
\label{eq:finpot}
\end{equation}
and the three-dimensional field components are:
\begin{equation}
\begin{split}
& B_r(r,\theta,z) = {\mathcal Re} \left \{ \sum_{n=0}^\infty  e^{i
(n+1)\theta} \sum_{k=0}^\infty {\frac{(-1)^k (n+1+2k)
(n+1)!}{2^{2k}(n+1+k)!k!}}  {\cal G}^{[2k]}_{n+1}(z) \, r^{n+2k}
\right \}  \\ & B_{\theta}(r,\theta,z) = -{\mathcal Im} \left \{
\sum_{n=0}^\infty e^{i (n+1)\theta} \sum_{k=0}^\infty {\frac{(-1)^k
(n+1)!(n+1)}{2^{2k}(n+1+k)!k!}} {\cal G}^{[2k]}_{n+1}(z) \, r^{n+2k}
\right \} \\ & B_z(r,\theta,z) =  {\mathcal Re} \left \{
\sum_{n=0}^\infty e^{i (n+1)\theta} \sum_{k=0}^\infty {\frac{(-1)^k
(n+1)!}{2^{2k}(n+1+k)!k!}} {\cal G}^{[2k+1]}_{n+1}(z) \, r^{n+1+2k}
\right \}
\end{split} \;\;. 
\label{eq:cylcomp}
\end{equation}
The coefficients ${\cal G}_{n+1}\equiv{\cal G}_{n+1,n+1}$ can be
related with the usual multipole coefficients, through
Eqs.~(\ref{eq:mult}). First, we write the scalar magnetic potential in
Cartesian coordinates:
\begin{equation}
\Phi(x,y,z)= {\mathcal Re} \left \{ \sum_{n=0}^\infty\sum_{k=0}^\infty
{\frac{(-1)^k (n+1)!}{2^{2k}(n+1+k)!k!}}  {\cal G}^{[2k]}_{n+1}(z) \,
(x+iy)^{n+1} (x^2+y^2)^{2k} \right \} \;\;.
\label{eq:potcart}
\end{equation}
The magnetic field components are computed by the gradient of the
potential (\ref{eq:potcart}):
\begin{equation}
\begin{split}
B_x(x,y,z) = {\mathcal Re} \Biggl\{  \sum_{n,k=0}^\infty
{\frac{(-1)^k (n+1)!}{2^{2k}(n+1+k)!k!}} &  (x^2+y^2)^{k-1}
{(x+iy)}^{n+1}\quad \times\\ & \left[ (n+1+2k)x -i(n+1)y \right] {\cal
G}^{[2k]}_{n+1}(z) \Biggr\} \\ B_y(x,y,z) = {\mathcal Im} \Biggl\{
\sum_{n,k=0}^\infty {\frac{(-1)^k (n+1)!}{2^{2k}(n+1+k)!k!}}&
(x^2+y^2)^{k-1} (x+iy)^{n+1} \times\\ & \left[
-(n+1)x+i(n+1+2k)y\right]  {\cal G}^{[2k]}_{n+1}(z) \Biggr\} \\
B_z(x,y,z) =  {\mathcal Re} \Biggl\{ \sum_{n,k=0}^\infty {\frac{(-1)^k
(n+1)!}{2^{2k}(n+1+k)!k!}} &  \, (x+iy)^{n+1} (x^2+y^2)^{2k}{\cal
G}^{[2k+1]}_{n+1}(z) \Biggr\}
\end{split}
\;\;.
\label{eq:cartcomp}
\end{equation}
Using Eqs.~(\ref{eq:mult}), we get:
\begin{equation}
\begin{split}
b_n(z) = & - (n+1)! \; {\mathcal Im} \{{\cal G}_{n+1}(z)\}  - n!
\sum_{k=1}^{n/2} \frac{(-1)^k(n+1-2k)(n+1-2k)!}{2^{2k}(n+1+k)!k!}
{\mathcal Im} \{{\cal G}^{[2k]}_{n+1-2k}(z)\} \\ a_n(z) = & \quad\;
(n+1)!  \; {\mathcal Re} \{{\cal G}_{n+1}(z)\}  + n! \sum_{k=1}^{n/2}
\frac{(-1)^k(n+1-4k)(n+1-2k)!}{2^{2k}(n+1+k)!k!}  {\mathcal Re}
\{{\cal G}^{[2k]}_{n+1-2k}(z)\}
\end{split}\;\; ,
\label{eq:multcyn}
\end{equation}
where  the upper limit of both series is the integer part of
$n/2$. Thus, in the absence of longitudinal dependence of the field,
the normal and skew multipole coefficients are just scalar multiples
of the imaginary and real part of ${\cal G}_{n+1}(z)$. On the other
hand, the situation is more complicated in the case of 3D fields. By
inverting the series (\ref{eq:multcyn}), we have:
\begin{equation}
\begin{split}
{\mathcal Im} \{{\cal G}_{n+1}(z)\} = & - \frac{1}{n!}
 \sum_{k=0}^{n/2} {\cal R}^{nor}_{n,k} b^{[2k]}_{n-2k}(z)  \\
 {\mathcal Re} \{{\cal G}_{n+1}(z)\} = &  \quad\; \frac{1}{n!}
 \sum_{k=0}^{n/2}  {\cal R}^{sk}_{n,k}  a^{[2k]}_{n-2k}(z)
\end{split}\;\; ,
\label{eq:cynmult}
\end{equation}
where the coefficients ${\cal R}^{sk}_{n,k}$ and ${\cal
R}^{nor}_{n,k}$ can be computed order by order by the $j+1$ relations
\begin{equation}
\begin{array}{lll}
{\cal R}^{nor}_{n,0}= & {\displaystyle \frac{1}{(n+1)} }
 &{\displaystyle  \quad , \quad
 \sum_{k=0}^{j}\frac{(-1)^k(n+1-2k)(n+1-2k)!}{2^{2k}(n+1+k)!k!}  {\cal
 R}^{nor}_{n-2k,j-k} = 0 }\\ {\cal R}^{sk}_{n,0} = & {\displaystyle
 \frac{1}{(n+1)} } &{\displaystyle  \quad , \quad
 \sum_{k=0}^{j}\frac{(-1)^k(n+1-4k)(n+1-2k)!}{2^{2k}(n+1+k)!k!}  {\cal
 R}^{sk}_{n-2k,j-k} = 0 }
\end{array}\;\; ,
\label{eq:coeff}
\end{equation}
and $j$ runs from 1 to the integer part of $n/2$. Using the last
relations, the scalar potential and the magnetic field can be
expressed as a function of the usual multipole coefficients.  By
expanding the complex polynomials in the expression of the magnetic
field components, one recovers the expansions of the magnetic
fields~(\ref{eq:fieldmgen}) in Cartesian coordinates.

\section{EVALUATION OF R.M.S. END DEFLECTIONS}
In order to evaluate the r.m.s. deflection caused by a magnet end, we
start from the expressions~\eqref{eq:fringmult} by splitting the
product inside the brackets:
\begin{equation}
\begin{array}{lrl}
\Delta p^f_{x}  \approx - & {\displaystyle  {\frac{e
\overline{b_n}}{4(n+1)!}}} \Bigl[ &{\mathcal Re}\left\{
(x+iy)^n\right\} \left[(n+1)xx' +(n-1)yy'\right] \\ & +& {\mathcal
Im}\left\{ (x+iy)^n\right\} \left[-(n+3)xy'+(n+1)x'y)\right] \Bigr] \\
& & \\ \Delta p^f_{y}  \approx & {\displaystyle  {\frac{e
\overline{b_n}}{4(n+1)!}}} \Bigl[ &{\mathcal Re}\left\{
(x+iy)^n\right\} \left[(n+1)xy' - (n+3)x'y\right] \\ & +&  {\mathcal
Im}\left\{ (x+iy)^n\right\} \left[(n-1)xx'+ (n+1)yy')\right] \Bigr ]
\end{array}
\label{eq:fringmultsplit}
\;\;.
\end{equation}
The total r.m.s. transverse momentum kick imparted by the fringe field
is $(\Delta p^f_{\perp})_{\text rms} = \sqrt{\langle (\Delta
p^f_{x})^2 \rangle + \langle (\Delta p^f_{y})^2 \rangle}$, where the
operator $\langle . \rangle$ denotes the average over the angle
variables. An equivalent expression stands for the deflection due to
the body part of the field. The $\langle . \rangle$ operator is
linear, we can first compute the sum of squares of the momentum kicks
and then proceed to their averaging. Thus, we have:
\begin{equation}
\begin{array}{r}
(\Delta p^f_{\perp})_{\text rms}  \approx {\displaystyle  \frac{e
\overline{b_n}}{4(n+1)!}  \Biggl[ \langle f_1 \left({\mathcal
Re}\left\{ (x+iy)^n \right\}\right)^2  +  f_2  \left({\mathcal
Im}\left\{ (x+iy)^n \right\}\right)^2 }\\ {\displaystyle +2 f_3
{\mathcal Re}\left\{ (x+iy)^n \right\}  {\mathcal Im}\left\{ (x+iy)^n
\right\} \rangle \Biggr]^{1/2}} \\ \\ (\Delta p^b_{\perp})_{\text rms}
\approx {\displaystyle  \frac{ e  \overline{b_n} L_{\text{eff}}}{n!}
\left[ \left\langle  ( {\mathcal Re} \left\{ (x+iy)^n \right\} )^2 +
( {\mathcal Im} \left\{ (x+iy)^n \right\} )^2   \right\rangle
\right]^{1/2} }
\end{array}
\label{eq:totrwo}
\;\;,
\end{equation}
where $f_1$, $f_2$ and $f_3$ are:
\begin{equation}
\begin{array}{lll}
f_1 & = &
(n+1)^2x^2({x'}^2+{y'}^2)+y^2\left[(n+3)^2{x'}^2+(n-1)^2{y'}^2\right]
- 8 (n+1) xx'yy' \\ & &\\ f_2 & = &
x^2\left[(n-1)^2{x'}^2+(n+3)^2{y'}^2\right]+(n+1)^2y^2({x'}^2+{y'}^2)
- 8 (n+1) xx'yy' \\ & &\\ f_3 & = & 4\left[ - (n+1) (x^2+y^2)x'y' +
xy({x'}^2+{y'}^2)\right]
\end{array}
\label{eq:efs}
\;\;.
\end{equation}
We have the following relations for the real and imaginary part of
$(x+iy)^n$:
\begin{equation}
\begin{array}{ll}
{\mathcal Re} \left\{(x+iy)^n \right\}= & {\displaystyle
\sum_{l=0}^{[n/2]} (-1)^l \binom{n}{2l} x^{n-2l} y^{2l}} \\ &\\
{\mathcal Im} \left\{(x+iy)^n \right\}= & {\displaystyle
\sum_{l=0}^{[(n-1)/2]} (-1)^l \binom{n}{2l+1} x^{n-2l-1} y^{2l+1} }
\end{array}
\label{eq:realim}
\;\;,
\end{equation}
and thus:
\begin{equation}
\begin{array}{rl}
\left({\mathcal Re} \left\{(x+iy)^n \right\}\right)^2 &=
{\displaystyle   \frac{1}{2} \left[\left(x^2+y^2\right)^n + {\mathcal
Re}\left\{(x+iy)^{2n}\right\} \right] }\\ &= {\displaystyle
\frac{1}{2} \sum_{l=0}^{n}  \left[\binom{n}{l} +
(-1)^l\binom{2n}{2l}\right] x^{2n-2l} y^{2l}} \\ \\ \left({\mathcal
Im} \left\{(x+iy)^n \right\}\right)^2 &= {\displaystyle   \frac{1}{2}
\left[\left(x^2+y^2\right)^n - {\mathcal Re}\left\{(x+iy)^{2n}\right\}
\right] }\\ &= {\displaystyle \frac{1}{2}  \sum_{l=0}^{n}
\left[\binom{n}{l} - (-1)^l\binom{2n}{2l}\right] x^{2n-2l} y^{2l}} \\
& \\ {\mathcal Re} \left\{(x+iy)^n \right\} {\mathcal Im}
\left\{(x+iy)^n \right\} &= {\displaystyle  \frac{1}{2} {\mathcal
Im}\left\{(x+iy)^{2n} \right\} }\\ & = {\displaystyle \frac{1}{2}
\sum_{l=0}^{n}  (-1)^l\binom{2n}{2l+1} x^{2n-2l-1} y^{2l+1}}
\end{array} \;,
\end{equation}
where the upper limit of the last sum is taken to be $l=n$ for
uniformity in the equations, instead of the last non-zero term for
which $l=n-1$. Finally, it is straightforward to show that
\begin{equation}
\left({\mathcal Re} \left\{(x+iy)^n \right\}\right)^2 +
\left({\mathcal Im} \left\{(x+iy)^n \right\}\right)^2 = (x^2+y^2)^n =
\sum_{l=0}^{n} \binom{n}{l} x^{2n-2l} y^{2l} \;\;.
\end{equation}
After expanding the products in Eq.~(\ref{eq:totrwo}) and collecting
the terms of equal power in the transverse variables, we have that the
transverse kicks can be written in the following form:
\begin{equation}
\begin{array}{lll}
(\Delta p^f_{\perp})_{\text rms}  & \approx & {\displaystyle  \frac{e
\overline{b_n}}{4(n+1)!}  \left[  \sum_{l=0}^{n} ( \Omega_1 + \Omega_2
+ \Omega_3 + \Omega_4 + \Omega_5 + \Omega_6) \right]^{1/2}}  \\ & & \\
(\Delta p^b_{\perp})_{\text rms}  & \approx & {\displaystyle  \frac{ e
\overline{b_n} L_{\text{eff}}}{n!}  \left[ \sum_{l=0}^{n} \binom{n}{l}
\left\langle x^{2n-2l} \right\rangle \left\langle y^{2l} \right\rangle
\right]^{1/2} }
\end{array}
\label{eq:newkick}
\;\;,
\end{equation}
where the $\Omega_k$'s are
\begin{equation}
\begin{array}{lll}
\Omega_1 & = & {\displaystyle \left(
\omega_1(n,l)+\omega_2(n,l)\right)  \langle x^{2n-2l+2}{x'}^2 \rangle
\langle y^{2l} \rangle }\\ & & \\ \Omega_2 & = & {\displaystyle
\left(\omega_3(n,l)+\omega_4(n,l) \right) \langle x^{2n-2l}{x'}^2
\rangle \langle y^{2l+2} \rangle }\\ & & \\ \Omega_3 & = &
{\displaystyle \left(\omega_3(n,l)+\omega_5(n,l) \right) \langle
x^{2n-2l+2} \rangle \langle y^{2l}{y'}^2 \rangle }\\ & & \\ \Omega_4 &
= & {\displaystyle \left( \omega_1(n,l)+\omega_6(n,l) \right) \langle
x^{2n-2l} \rangle \langle y^{2l+2}{y'}^2 \rangle }\\ & & \\ \Omega_5 &
= &  {\displaystyle \omega_7(n,l) \langle x^{2n-2l-1}{x'} \rangle
\langle y^{2l+3}{y'} \rangle} \\  & & \\ \Omega_6 & = & {\displaystyle
\left( \omega_7(n,l)+\omega_8(n,l)\right)  \langle x^{2n-2l+1}{x'}
\rangle \langle y^{2l+1}{y'} \rangle}
\end{array}
\;\;,
\label{eq:omegas}
\end{equation}
with the coefficients $\omega_k$'s:
\begin{equation}
\begin{array}{llllll}
\omega_1(n,l) & = & {\displaystyle \left( n^2+1 \right)\binom{n}{l} }
& \omega_2(n,l) & = & {\displaystyle 2n (-1)^l \binom{2n}{2l}  }\\
\omega_3(n,l) & = & {\displaystyle  \left( n^2+4n+5
\right)\binom{n}{l}  }  & \omega_4(n,l) & = & {\displaystyle
\frac{2(5n+2ln+2) (-1)^l}{2l+1} \binom{2n}{2l}  }\\ \omega_5(n,l) & =
& {\displaystyle - 2\left( n+2\right) (-1)^l \binom{2n}{2l} } &
\omega_6(n,l) & = & {\displaystyle  \frac{- 2l(2n+1)
(-1)^l}{2l+1}\binom{2n}{2l}  }\\ \omega_7(n,l) & = &   {\displaystyle
\frac{-8 (n+1)(n-l)(-1)^l}{2l+1}\binom{2n}{2l}  }  &   \omega_8(n,l) &
= & {\displaystyle -8  \left(n+1\right) \binom{n}{l} }
\end{array}
\;\;.
\label{eq:coefomegas}
\end{equation}
In order to proceed to the averaging of the transverse variables, we
write them in the standard form
\begin{equation}
\{x,y\} = \sqrt{\epsilon_{x,y}\beta_{x,y}}\ C_{x,y},\quad \{x',y'\} =
\sqrt{\frac{\epsilon_{x,y}}{\beta_{x,y}}}(S_{x,y}+\alpha_{x,y}
C_{x,y}) \;\;,
\label{eq:emitang}
\end{equation}
where $\epsilon_{x,y}$ are the transverse emittance associated with
the corresponding phase space dimension, $\beta_{x,y}$, $\alpha_{x,y}$
are the usual beta and alpha functions and  $C_q,S_q$ stand for
$\cos{\phi_q},\sin{\phi_q}$, respectively. Using the above relations
and averaging over the angle variables $\phi_q$ one can show that:
\begin{equation}
\begin{array}{ll}
\langle q^{2m} \rangle & =  {\displaystyle  \binom{2m}{m}
\frac{\beta_q^m\epsilon_q^{m}}{2^{2m}} } \\ & \\ \langle q^{2m} {q'}^2
\rangle & =  {\displaystyle \binom{2m}{m}
\frac{\left[1+(2m+1)\alpha_q^2\right]
\beta_q^{m-1}\epsilon_q^{m+1}}{2^{2m+1}(m+1)} }
\\ &  \\ \langle q^{2m+1} q' \rangle & = {\displaystyle
\binom{2(m+1)}{m+1}
\frac{\alpha_q\beta_q^m\epsilon_q^{m+1}}{2^{2m+2}} }
\end{array}
\label{eq:averaging}
\;\;.
\end{equation}
Then, the $\Omega_k$'s become:
\begin{equation}
\begin{array}{ll}
\Omega_1 & =  {\displaystyle \left(\omega_1(n,l) +
\omega_2(n,l)\right) \binom{2(n-l)}{n-l} \binom{2l}{l}
{\frac{(2n-2l+1)[1+(2n-2l+3)\alpha_x^2] \beta_x^{n-l}\beta_y^{l}
\epsilon_x^{n-l+2}\epsilon_y^l}{2^{2n+2}(n-l+1)(n-l+2)}} }\\ &  \\
\Omega_2 & =  {\displaystyle \left(\omega_3(n,l) +
\omega_4(n,l)\right) \binom{2(n-l)}{n-l} \binom{2l}{l}
{\frac{(2l+1)[1+(2n-2l+1)\alpha_x^2] \beta_x^{n-l-1}\beta_y^{l+1}
\epsilon_x^{n-l+1}\epsilon_y^{l+1}}{2^{2n+2}(n-l+1)(l+1)}} }\\ &  \\
\Omega_3 & =  {\displaystyle \left(\omega_3(n,l) +
\omega_5(n,l)\right) \binom{2(n-l)}{n-l} \binom{2l}{l}
{\frac{(2n-2l+1)[1+(2l+1)\alpha_y^2] \beta_x^{n-l+1}\beta_y^{l-1}
\epsilon_x^{n-l+1}\epsilon_y^{l+1}}{2^{2n+2}(n-l+1)(l+1)}} }\\ &  \\
\Omega_4 & =  {\displaystyle \left(\omega_1(n,l) +
\omega_6(n,l)\right) \binom{2(n-l)}{n-l} \binom{2l}{l}
{\frac{(2l+1)[1+(2l+3)\alpha_y^2] \beta_x^{n-l}\beta_y^{l}
\epsilon_x^{n-l}\epsilon_y^{l+2}}{2^{2n+2}(l+1)(l+2)}} }\\ &  \\
\Omega_5 & =  {\displaystyle \omega_7(n,l) \binom{2(n-l)}{n-l}
\binom{2l}{l}  {\frac{(2l+1)(2l+3)\alpha_x\alpha_y
\beta_x^{n-l-1}\beta_y^{l+1}
\epsilon_x^{n-l}\epsilon_y^{l+2}}{2^{2n+2}(l+1)(l+2)}} }\\ &  \\
\Omega_6 & =  {\displaystyle \left(\omega_7(n,l) +
\omega_8(n,l)\right) \binom{2(n-l)}{n-l} \binom{2l}{l}
{\frac{(2n-2l+1)(2l+1)\alpha_x\alpha_y \beta_x^{n-l}\beta_y^{l}
\epsilon_x^{n-l+1}\epsilon_y^{l+1}}{2^{2n+2}(n-l+1)(l+1)}} }
\end{array}.
\label{eq:omegasav}
\end{equation}
After collecting terms of equal emittances, the r.m.s. transverse
momentum kicks can be expressed as:
\begin{equation}
\begin{array}{l}
(\Delta p^f_{\perp})_{\text rms}  \approx  {\displaystyle  \frac{e
\overline{b_n}}{2^{n+3}(n+1)!}  \left[  \sum_{l=0}^{n}
\binom{2(n-l)}{n-l} \binom{2l}{l}
\beta_x^{n-l}\beta_y^{l}\epsilon_x^{n-l}\epsilon_y^{l}
\sum_{m=0}^{2}g_{n,l,m}(\alpha_{x,y},\beta_{x,y})\epsilon_x^m
\epsilon_y^{2-m} \right]^{1/2}}  \\ \\ (\Delta p^b_{\perp})_{\text
rms}   \approx  {\displaystyle  \frac{ e  \overline{b_n}
L_{\text{eff}}}{2^nn!}  \left[ \sum_{l=0}^{n}  \binom{n}{l}
\binom{2(n-l)}{n-l} \binom{2l}{l}  \overline{\beta_x^{n-l}}
\overline{\beta_y^{l}}  \epsilon_x^{n-l} \epsilon_y^{l} \right]^{1/2} }
\end{array}
\label{eq:finkick}
\hskip -15pt ,
\end{equation}
where the bars on the $\beta$'s denote their average values over the
body of the magnet. The coefficients $g_{n,l,m}$, given by
\begin{equation}
\begin{array}{lll}
g_{n,l,0}(\alpha_{x,y},\beta_{x,y}) & = &  {\displaystyle
\frac{\left[(n^2+1)(2l+1)\binom{n}{l}-
2l(2n+1)(-1)^l\binom{2n}{2l}\right]
[1+(2l+3)\alpha_y^2]}{(l+1)(l+2)}}\\ & &\\ & & {\displaystyle -\frac{8
(n+1)(n-l)(2l+3)(-1)^l\binom{2n}{2l}
\alpha_x\alpha_y\beta_y}{\beta_x(l+1)(l+2)}}\\ & &\\
g_{n,l,1}(\alpha_{x,y},\beta_{x,y}) & = & {\displaystyle
\frac{\left[(n^2+4n+5)(2l+1)\binom{n}{l}+2(5n+2ln+2)(-1)^l\binom{2n}{2l}
\right][1+(2n-2l+1)\alpha_x^2]\beta_y}{\beta_x (n-l+1)(l+1)}}\\ & &\\
&  &{\displaystyle + \frac{\left[(n^2+4n+5)\binom{n}{l}
-2(n+2)(-1)^l\binom{2n}{2l}\right]
(2n-2l+1)[1+(2l+1)\alpha_y^2]\beta_x}{\beta_y(n-l+1)(l+1)}}\\ & &\\ &
&{\displaystyle -\frac{8 (n+1)\left[(2l+1)\binom{n}{l}+(n-l)(-1)^l
\binom{2n}{2l} \right] (2n-2l+1) \alpha_x\alpha_y}{(n-l+1)(l+1)}}\\ &
&\\ g_{n,l,2}(\alpha_{x,y},\beta_{x,y}) & = &  {\displaystyle
\frac{\left[\left( n^2+1 \right)\binom{n}{l}+ 2n (-1)^l \binom{2n}{2l}
\right](2n-2l+1) [1+(2n-2l+3)\alpha_x^2]}{(n-l+1)(n-l+2)}}\\
\end{array}
\label{eq:ges}
,
\end{equation}
depend on the twiss functions $\alpha_{x,y}$, $\beta_{x,y}$ and on the
multipole order $n$.  One may  note that r.m.s. transverse momentum
kick of the fringe is represented by the square root of a polynomial
of order $n+2$ in the transverse emittances $\epsilon_x$ and
$\epsilon_y$ as compared to the square root of a polynomial of order
$n$ representing the body contribution (see also~\cite{Forest}). Thus,
their ratio should be proportional to the transverse emittance. This
scaling law is indeed exact for the case of the ``dipole'' and
``quadrupole''. For higher order ``multipoles'', it is exact for flat
and round beams (Sec. IV).

\section{Acknowledgements}
The authors would like to thank A.~Jain for useful suggestions
regarding the magnetic field expansions, E.~Keil for his criticism in
an early version of this work and R.~Baartman for many useful
comments and discussion. This work was performed under the auspices of
the U.S. Department of Energy.

\begin{table}
\caption{Parameters associated with the LHC and SNS magnets, whose
fringe-field figure of merit is evaluated in
Fig.~\ref{fig:fringeLHC}. When two numbers occur, they are associated
to the minimum and maximum value.} 
\begin{center}
\begin{tabular}{|l|r|c|c|c|c|c|}
\hline
Magnet & Number & $L_{\text{eff}}$ [m] & $\beta_{x,y}$ [m] &
$\overline{\beta_{x,y}}$ [m] & $|\alpha_{x,y}|$ [m] & $\epsilon_{x,y}$
[m~rad] \\   
\hline 
LHC Quadrupole Triplets & 16   & 5.5 -- 6.37 & 1055 -- 4463 & 1157 --
4401 & 1.1 -- 203.9 & 5.03 10$^{-10}$  \\  
LHC Arc Quadrupoles 	& 368  & 3.1 	     & 32 -- 178    & 32 --
176  & 0.5 -- 2.4   & 7.82 10$^{-9}$  \\ 
LHC Dipoles 		& 1104 & 14.3        & 28 -- 176    & 40 --
143  & 0.5 -- 2.6   & 7.82 10$^{-9}$  \\  
SNS Dipoles 		& 32   & 1.5         & 4 -- 8  & 6 & 1.1 --
1.9 & 4.8 10$^{-4}$ \\ 
SNS Quadrupoles 	& 52   & 0.5 -- 0.7  & 2 -- 28 & 2 -- 26 & 0 -- 8 & 4.8 10$^{-4}$ \\ 
\hline
\end{tabular}
\end{center}
\label{tab:param}
\end{table}

\end{document}